\documentclass[epj,referee]{svjour}
\usepackage{graphics}
\def\br{\vec{r}}

\def\bn{\vec{n}}

\def\bR{\vec{R}}

\def\bS{\vec{S}}

\def\brho{\vec{\rho}}
\def\bzeta{\vec{\zeta}}

\def\hC{{\hat C}}

\def\heps{{\hat \epsilon}}

\def\cW{{\cal W}}
\def\ccP{{\cal P}}
\def\cP{\Theta}

\def\lX{{\lambda X}}

\begin{document}
\title{Geometric partition functions of cellular systems: Explicit calculation of the entropy in two and three dimensions}
\author{Raphael Blumenfeld\inst{1,2} \and Sam F. Edwards\inst{2}
\thanks{RB is grateful to Prof N. Rivier for an illuminating discussion}
}                     
%
%
\institute{Earth Science and Engineering, Imperial College, London SW7 2BP, UK \and Biological and Soft Systems, Cavendish Laboratory, Madingley Road, Cambridge CB3 0HE, UK}
\date{Received: date / Revised version: date}
%
\abstract{
A method is proposed for the characterisation of the entropy of cellular structures, based on the compactivity concept for granular packings. Hamiltonian-like volume functions are constructed both in two and in three dimensions, enabling the identification of a phase space and making it possible to take account of geometrical correlations systematically. Case studies are presented for which explicit calculations of the mean vertex density and porosity fluctuations are given as functions of compactivity. The formalism applies equally well to two- and three-dimensional granular assemblies.
} 
\maketitle
\section{Introduction}
\label{intro}
Many foams and cellular systems appear to evolve into steady states of similar structures irrespective of the specific 
dynamics and of the lengthscales over which the dynamics take place. The lengthscales can differ by orders of magnitude
between systems \cite{FoamRev}. This prompted the use of statistical mechanical methods to analyse some of the characteristics of such structures \cite{FoamEntropy}. This approach has remained underdeveloped for several reasons:
(a) it has been unclear why equlibrium-based description should apply to these non-equilibrium systems; (b) it has been difficult to identify a phase space that makes it possible to either disentangle the geometrical correlations or at least enable to treat them systematically; (c) it has been unclear what should play the role of a Hamiltonian, which is at the foundation of the conventional statistical mechanical formalism; (d) the extension of the concept of temperature is far from obvious in these athermal systems.  

This paper proposes a formalism based on the concept of compactivity \cite{EdwardsEntropy}, which resolves all these difficulties. 
In the first part we adapt a basic formalism, originally proposed for granular systems, to planar cellular systems \cite{BlEd03}. 
A convenient volume function is proposed as the natural analog of the Hamiltonian in conventional thermodynamic systems. The volume function makes it possible to identify a phase space and a method is outlined to isolate and systematically analyse the correlations due to topological constraints. 
In the second part we use the insight gained from planar systems to construct the formalism for three-dimensional structures. The key step is the identification of an exact volume function which makes it possible to pinpoint the configurational phase space and evaluate its dimensionality. We illustrate the formalism with explicit calculations of several case studies.

The compactivity concept \cite{EdwardsEntropy} has been introduced in the context of granular packings to enable the use of statistical mechanical methods for the analysis of vibrated granular systems. Relying on observations that under specified vibrations granular assembliess settle into packings of reproducible densities, it was proposed that there is a steady state distribution of configurations. By identifying the configurations a statistical mechanical-like approach can be constructed.
However, thermal energy, which is relevant in conventional molecular systems, is irrelevant in granular systems. This led to the suggestion to replace the conventional Hamiltonian by a volume function $\cW$. The entropy of such systems is volumteric and it depends on the volume of the system $V$ and on the number of grains that an assembly of many grains can take. To move from configuration to configuration these athermal systems require external agitation. Such agitation has been quantified by a new parameter, compactivity, defined as $X = \partial V/\partial S$.

In cellular systems the situation apears to be different. Not only the dynamics by which configurations may change are different but also foams, for example, appear to be always in non-equilibrium, with bubbles ever growing in size. 
Nevertheless, it is often observed \cite{FoamEntropy} that foams and cellular systems converge to a scaling regime that can be regarded as a steady state. In foams the distributions of several structural properties (e.g. the volumes of cells) is stationary when the space dimension is scaled by an appropriate power of time, $r\to r'=r/t^x$. The existence of a steady state suggests that these out-of-equilibrium systems may be analysed using statistical mechanical methods \cite{FoamEntropy}. Following the same logic as in granular systems one can apply the concept of compactivity to foams and cellular systems. In this formalism the partition function of a canonical ensemble in $d$ dimensions is

\begin{figure}
\resizebox{0.5\textwidth}{!}
{
  \includegraphics{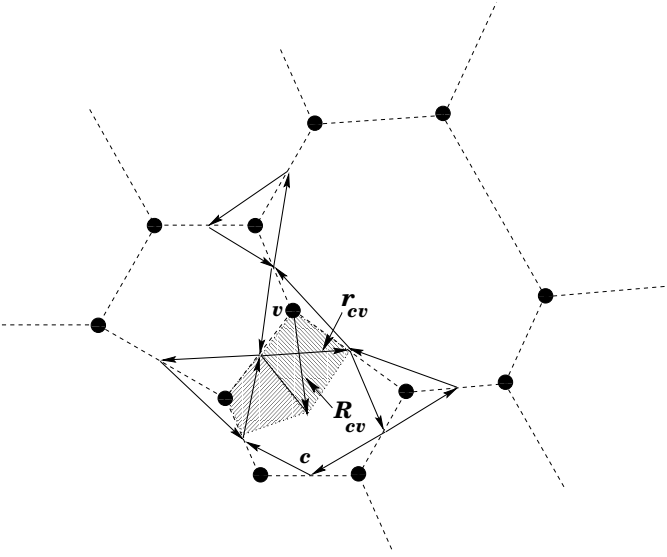}
}
\caption{The geometric construction around vertex $v$ in two dimensions. The vectors $\br_{cv}$ connect midpoints of cell walls clockwise around each vertex. The vector $\bR_{cv}$ extends from the centre of triangle $v$ to the centre of cell $c$. The antisymmetric part of each term in the tensor $\hC_v = \sum_l{\bR_{cv}\br_{cv}}$ describes the area of the quadrilateral of which $\bR_{cv}$ and $\br_{cv}$ are the diagonlas. This quadrilateral and a neighbouring one are shown shaded. The quadrilateral are exactly adjacent and therefore a sum over these terms give exactly the volume of the system.}
\label{fig:1}       
\end{figure}

\begin{equation}
Z = \int e^{-{{\cW_d(\{q_n\})}\over {\lX}}} \cP(\{q_n\}) \prod_n dq_n
\label{eq:Aiv}
\end{equation}  
where $\{q_n\}$ are the degrees of freedom that comprise the phase space and $\cP(\{q_n\})$ is the probability density of finding the system at a particular point in this space. This density is nontrivial due to the connectivity forced by the mechanical equilibrium. $\lambda$ is the analog of Boltzmann's constant and $\beta\equiv 1/\lX$ has dimensions of inverse volume. Since the term 'free volume' has other associations in the literature let us term here the analog of the free energy the free porosity, $Y = -\ln Z/\beta$. In direct analogy with thermodynamics, the (dimensionless) volumetric entropy is $S = \beta^2 \partial Y / \partial \beta$ and the mean porosity is $\langle V\rangle = Y + S / \beta$. 
Note that whether the ensemble is canonical or microcanonical depends on whether the volume function is constant across the ensemble. The particular choice of ensemble is not essential to the following analysis and we shall not concern ourselves with it any further. 

The main difficulties in the application of this formalism have been the identification of a volume function which is exactly additive when summed over the basic elements of the systems (cells or grains) and the isolation of a manageable phase space of independent degrees of freedom. In the following we focus on the canonical ensemble at a fixed, but large, number of cells. The extension to grand canonical ensembles where cells are continually created and destroyed is straightforward.

The main aim of this paper is to formulate a three-dimensional exact volume function in terms of local structural descriptors and to demonstrate its use. However, as will become clear below, an important step toward this goal is the adaptation of a recent formulation of an exact volume function in two-dimensions \cite{BlEd03}.

\section{Two dimensional structures}
\label{sec:1}

A planar dry foam comprises of $N\gg 1$ vertices, each of which connects three edges (trivalent structures). Vertices that connect more edges are rare in many common circumstances and we shall disregard those in this paper. Nevertheless, the formalism developed here is quite general and applies to non-trivalent foams as well. The edges can be either straight or curved; the structural characterisation described here works for both cases and for systems that comprise of a mixture of such edges. The edges partition the plane into $N/2$ polygonal (or polygonal-like in the case of curved edges) cells with an average of six edges per cell \cite{SixEdge} and altogether $3N/2$ edges. This neglects boundary corrections which are {\it at most} of order O($\sqrt{N}$).

Consider the following network \cite{Bl03}. Around each vertex draw a triangle whose corners are the midpoints of the edges that lead to it (see figure 1). We regard the sides of each triangle as vectors that circulate clockwise around the vertex. Consequently these vectors form closed loops around cells which circulate anticlockwise. A vector around vertex $v$ and within cell $c$ is indexed $\br_{cv}$. The vectors $\br_{cv}$ form a network that spans the system, the $\br$-network.
Let $\brho_v$ be the mean position (the centroid) of the corners of the triangle around $v$ and $\brho_c$ be the centroid of the midpoints of the edges surrounding cell $c$. 
We define a vector that extends from $\brho_v$ to $\brho_c$, $\bR_{cv} = \brho_c - \brho_v$.  The network of $\bR$ vectors is dual to the $\br$-network. Each $\br_{cv}$-$\bR_{cv}$ pair forms a quadrilateral of which they are the diagonals, such as the two shown shaded in figure 1. The shape and geometry of a quadrilateral is characterised by the outer product

\begin{equation} 
C_{cv}^{ij} =  r_{cv}^i R_{cv}^j 
\label{eq:Ai} 
\end{equation} 
where $i,j = x,y$ are Cartesian components. The tensor $\hC_v = \sum_c \hC_{cv}$, where the sum runs over the three cells surrounding vertex $v$, has been found to play a central role in the analysis of stress transmission both in cellular  systems \cite{Bl03} and in granular assemblies \cite{BaBl02}.  
The antisymmetric part of $\hC_{cv}$ is $\hC_{cv} - \hC_{cv}^T = A_{cv}\heps$ where $\matrix{\heps}={0 \,\ 1 \choose -1 \,0 }$ and $A_{cv}=\br_{cv}\times\bR_{cv}/2$ is the area of the quadtrilateral. 
The choice of the directions of $\br_{cv}$ and $\bR_{cv}$ ensures that all the areas $A_{cv}$ are pseudo-vectors that point in the same direction, into the page. This direction is taken to be positive. A key observation
is that the quadrilaterals perfectly cover the plane, $A_{sys} = \sum_{cv} A_{cv}$, providing an exact volume function. This volume function  resolves the difficulties in the application of the compactivity-based formalism which were mentioned in the introduction. The two-dimensional volume function is 

\begin{equation} 
\cW_2 = \frac{1}{2}{\rm Tr}\left[\sum_{cv} \hC_{cv}\cdot\heps^T\right] \ .
\label{eq:Aii} 
\end{equation} 
We now wish to determine the phase space, i.e. the independent degrees of freedom. The volumes $A_{cv}$ are expressed in terms of the  vectors $\br_{cv}$ and $\bR_{cv}$ and we note that there are altogether $3N$ vectors of each. But these are not all independent.  First, the self-duality between the $\bR$ and $\br$ networks means that all the $\bR$ vectors can be determined in terms of linear combinations of the $\br$ vectors. 
Specifically, every $\bR_{cv}$ can be written as a linear combination of exactly $(Z_c+Z_v-2)$ $\br$ vectors, where $Z_c$ is the number of edges of cell $c$ and $Z_v$ is the number of vectors around vertex $v$ (in trivalent cellular systems $Z_v=3$, but it is arbitrary in granular assemblies and in non-trivalent foams). To see this recall that $\bR_{cv}$ extends from the centroid of vertex $v$, $\brho_v=\frac{1}{Z_v}\sum_{c'=1}^{Z_v-1}(Z_v-c')\br_{c'v}$, to that of cell $c$, $\brho_c=\frac{1}{Z_c}\sum_{v'=1}^{Z_c-1}(Z_c-v')\br_{cv'}$. Therefore, $\bR_{cv} = \brho_c - \brho_v = \sum_{k=1}^{Z_c + Z_v - 2} a_n \br_n$ can be expressed as a {\it linear combination} of the vectors circulating triangle $v$ and cell $c$. 
An example is shown in figure 2 where $\bR_3$, which points to the centroid of a pentagonal cell, is a function of $5+3-2=6$ $\br$ vectors, 

\begin{figure}
\resizebox{0.5\textwidth}{!}
{
  \includegraphics{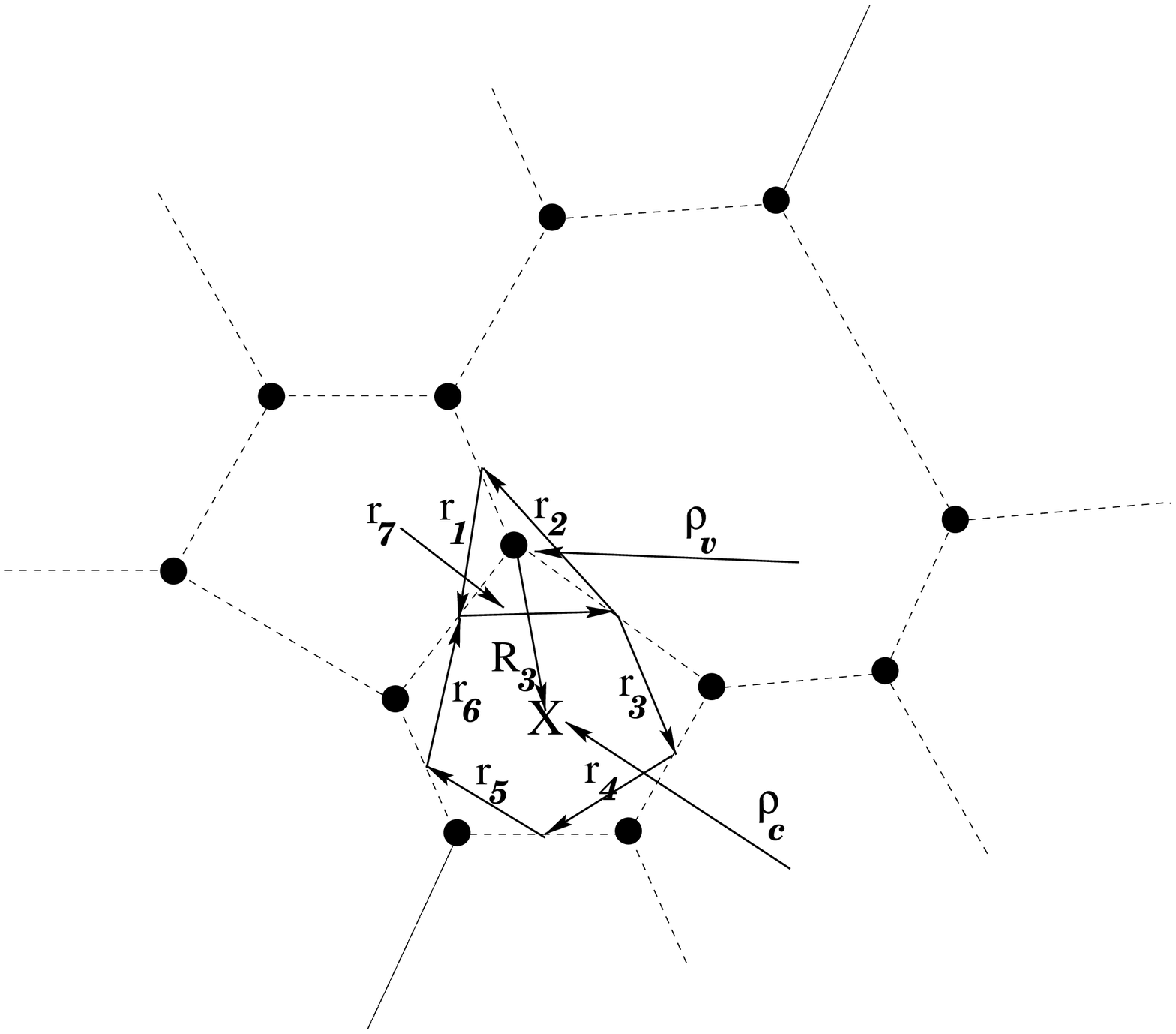}
}
\caption{The vector $\bR_3$ of the dual network is a linear combination of the contact network vectors $\br_n$ ($n=1-6$). Note that it does not depend on the vector $r_7$ and therefore $\bR_{cv}$ depends only on $(z_c + 1)$ $\br$ vectors. $\brho_c$ is the position vector of the centroid of the cell (marked by X). $\brho_v$ is the centroid of the triangle made by the vectors $\br_1$, $\br_2$ and $\br_7$. }
\label{fig:2}       
\end{figure}

\begin{equation} 
\bR_3 = \frac{1}{5}\left(4r_3 + 3r_4 + 2r_5 + r_6\right) - \frac{1}{3}\left(2r_2 + r_1\right)
\label{eq:Aiii} 
\end{equation} 
On average there are six edges around cells and therefore an $\bR$ vector depends, on average, on $\langle Z_v + Z_c - 2\rangle = 7$ $\br$ vectors.
Thus, there remain only the $3N$ $\br$ vectors to examine for independence. These are also inter-related due of the loops that they form. The inter-dependence is directly evaluated in terms of the number of {\it irreducible} loops, namely, the elementary loops in terms of which all other loops can be expressed. In a loop of $n$ vectors only $n-1$ are independent, so every irreducible loop provides one dependent vector. There are only two types of such loops: the triangles around the vertices, of which there are $N$, and the polygons around the cells, of which there are $N/2$. This means that of the $3N$ $\br$ vectors, $3N/2$ are  dependent. The independent degrees of freedom are the components of the remaining $3N/2$ vectors and therefore the phase space is $3N$-dimensional.

We now note a surprising coincidence: the number of independent degrees of freedom is equal to the number of quadrilaterals! This, together with the fact that the sum over the areas of the quadrilaterals is the volume function, suggests that the quadrilaterals are the actual elementary `quasi-particles' of the system, not the vertices or the cells (nor the grains in the analogue context). It is convenient to express the phase space in terms of their areas, $A_{cv}\equiv A_q$. 
We term these elementary quasi-particles `Quadrons'. In terms of the quadrons the partition function is 

\begin{equation}
Z = \int \prod_q \left[ e^{-\beta A_q } dA_q \right] \cP(\{A_q\})
\label{eq:Ci}
\end{equation}
and the function $\cP(\{A_q\})$ can be interpreted as a density of volumes (the analogue of the density of states in conventional statistical mecanics). It should be noted that cellular and granular systems support different limits of integration in (\ref{eq:Ci}). In the former the lower limit on $A_q$ may be as vanishingly small as the thickness of a cell wall and the upper limit, subject to the constraint that the total number of vertices remains $N$, may be almost as large as the entire system. In granular packings $A_q$ depends strongly on the grain size distribution and is bounded by finite upper and lower values due to the fact that grains cannot inter-penetrate. 

To illustrate the method, let us consider very large skeletal cellular systems (i.e. of vanishing cell wall thickness) and assume for simplicity that the quadrons are independent, $\cP(\{A_q\})=\pi_{q=1}^{3N}\ccP(A_q)$. We are not aware of any first-principles calculation of $\cP(\{A_q\})$ for disordered cellular structures and therefore we shall postulate several simple forms and analyse briefly the insight that they offer. First, let us consider the case of a uniform distribution between lower and upper volumes $A_0$ and $A_1$, repectively, 

\begin{equation}
\ccP(A_q) = \cases{\frac{2}{\delta A} &if $A_0 < B_q < A_1$; \cr
                0 &otherwise \ , \cr}
\label{eq:Cii}
\end{equation}
where $\delta A\equiv(A_1-A_0)/2$. In dynamically evolving systems $\delta A$ and $\bar{A} \equiv (A_1+A_0)/2$ would scale with time. Substituting this form into the partition function gives

\begin{eqnarray}
Z & = & \left[ \int_{A_0}^{A_1} e^{-\beta A} dA \right]^{3N} = 
\left[ \frac{\beta e^{-\beta \bar{A}} {\rm sinh}(\beta \delta A)}{\beta \delta A}\right]^{3N} \nonumber \\ 
& {\rm and} & \nonumber \\ 
S & = & 3N \left[ 1 + {\rm ln}\frac{{\rm sinh}(\beta \delta A)}{\beta \delta A} - \frac{\beta \delta A}{ {\rm tanh}( \beta \delta A )} \right] \ .
\label{eq:Ciii}
\end{eqnarray}
The mean volume (or porosity) of this system per quadron is then 
$$\langle V \rangle = \bar{A} + 1/\beta - \frac{\delta A}{{\rm tanh}(\beta\delta A)}$$ 
and the mean porosity fluctuations is 
$$\langle \delta V^2 \rangle = \beta^{-2} - \frac{(\delta A)^2}{{\rm sinh}^2(\beta\delta A)} \ .$$ 

As a second example consider a binary mixture of two types of independent quadrons, $A_0$ and $A_1$, occuring at respective probabilities $p$ and $1-p$

\begin{equation}
\ccP(A_q) = p\delta(A_q - A_0) + (1-p)\delta(A_q - A_1) \ .
\label{eq:Civ}
\end{equation}
The one quadron partition function is found to be

\begin{equation}
z = 2\sqrt{p(1-p)}e^{-\beta\bar{A}}{\rm cosh}\left(\beta \delta A + u\right)  \ ,
\label{eq:Cv}
\end{equation}
where $u = {\rm ln}[p/(1-p)]$ and $\bar{A}$ and $\delta A$ are defined as in the previous case. The entropy per quadron can be calculated from this expresion using $s\equiv S/N_{dof}={\rm ln}z-\beta\partial{\rm ln}z/\partial\beta$ and the mean porosity (volume) per quadron is 
$$\langle V \rangle = \bar{A} - \delta A{\rm tanh}(\beta\delta A + u ) \ .$$ 
The mean of the porosity fluctuations per quadron is
$$\langle \delta V^2 \rangle = \left(\frac{\delta A}{{\rm cosh}(\beta \delta A + u )}\right)^2 \ .$$
It is worth noting that as a function of the agitation the system can occupy volumes between $\bar{A} + \delta A$ and $\bar{A} + \delta A {\rm tanh}(u)$ and that this case is the exact analogue of the Bragg-Williams description of alloys \cite{BrWi}.

For a third example we are prompted by observations, e.g. in soap froths \cite{FoamRev}\cite{FoamEntropy}, to consider a system with a broad distribution of cell sizes. Such distributions have generically algebraic tails of the form

\begin{equation}
\ccP(A_q) = CA_q^{x-1} \ \ \ \ ;\ \ \ 0 < A_q < \alpha \ ,
\label{eq:Cvi}
\end{equation}
where $0<x<1$ and $C$ is a normalisation constant. The partition function is straightforward to compute 

\begin{equation}
Z = \left[ C\beta^{-x} \gamma(x,\beta\alpha) \right]^{3N} \ ,
\label{eq:Cvii}
\end{equation}
where $\gamma(x,y)$ is the incomplete gamma-function \cite{GrRy}. As in the previous cases, the entropy can be obtained directly from this expression. The mean porosity per quadron is 
$$\langle V \rangle = \frac{\gamma(x+1,\beta\alpha)}{\beta \gamma(x,\beta\alpha)} $$
and the mean porosity fluctuations is
$$\langle \delta V^2 \rangle = \frac{\gamma(x,\beta\alpha)\gamma(x+2,\beta\alpha)-\gamma(x+1,\beta\alpha)^2} {\beta^2\gamma(x,\beta\alpha)^2} \ .$$

\begin{figure}
\resizebox{0.5\textwidth}{!}
{
  \includegraphics{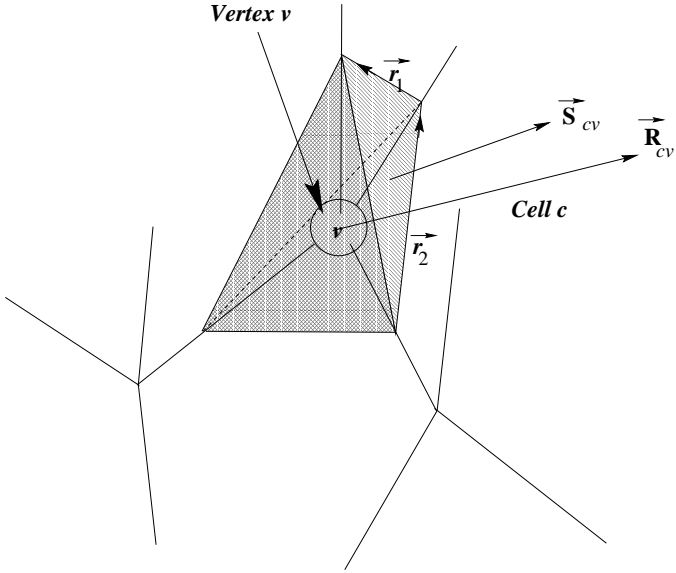}
}
\caption{The geometric construction around vertex $v$. Vectors such as $\br_1$ and $\br_2$ connect midpoints of struts around each vertex to form a tetrahedron. The vectors $\bR_{cv}$ points from the center of tetrahedron $v$ to that of its neighbouring cell $c$. $\bS_{cv}=\br_1\times\br_2/2$ is the directed area of the face of the tetrahedron between $v$ and $c$.}
\label{fig:3}       
\end{figure}

\section{Three dimensional structures}
\label{sec:2}

We can use the insight from the two-dimensional analysis as a basis for the formulation of the three-dimensional case. In three dimensions systems may be either close-cell, in which case the cells are enclosed by material membranes (cell walls), or open-cell, in which case vertices are connected by struts of given thicknesses and the cells are interconnected. A sponge is an example of an open-cell structure. The formalism presented here is based on the structure of the {\it skeleton} (see below) and is therefore a description of its {\it topology}. As such it applies equally well to both types of structures. Nevertheless, it is convenient to present the following discussion in the context of open-cell structures.

Most common three-dimensional cellular structures are quadrivalent, namely, when each vertex connects to exactly four neighbour vertices.  
The skeleton of the structure is obtained by extending lines between neighbouring vertices. The lines can be either straight or curved and represent the struts thinned down to vanishing widths. The skeletal struts meeting at a vertex form angles that are presumed in the following to be random. If they are not (as may happen, e.g. due to dynamics processes dominated by surface tensions) then the configuration space is spanned by fewer independent degrees of freedom. For simplicity, we shall not discuss this class of materials here, but the analysis can be extended to it straightforwardly.

Consider the following construction. Connect the midpoints of the struts around every vertex by imaginary straight lines, which we term edges. The edges enclosing each vertex form a tetrahedron, as shown in figure 3.
Tetrahedra around two neighbouring vertices make contact at the midpoint of the strut between them. The tetrahedra form an interconnected framework that spans the system which we term, following intuition from granular systems, the contact network. Every tetrahedron exposes one triangular face to each of the four cells around it. Denoting a vertex by $v$ and a cell by $c$, every such triangular face can be indexed uniquely by `$cv$'. The triangular face that tetrahedron $v$ exposes to cell $c$ is characterised by $\bS_{cv}=S_{cv}\bn_{cv}$, where $S_{cv}$ is the area of the triangle and $\bn_{cv}$ a unit normal to its plane, which points away from the vertex and into cell $c$. The edges that make this triangular face are regarded as vectors circulating the triangle in the clockwise direction when viewed from the vertex, for example vectors $\br_1$ and $\br_2$ in figure 3. 

The centroid of the triangle $S_{cv}$, $\bzeta_{cv}$, is defined as the mean position of its three corners. We also define the centroid of tetrahedron $v$, $\brho_v$ as the mean position of its four corners and the centroid of cell $c$, $\brho_c$ as the mean position of the midpoints of the struts that surround this cell. Note that while $\brho_v$ is expected to be in the vicinity of vertex $v$ it need not coincide with it. Nevertheless, for simplicity, we shall refer to the point $\brho_v$ from now on as vertex $v$ unless otherwise stated.

Let us inspect the contact network from inside cell $c$. From such a point the `surface' of the cell consists of the triangular faces of the tertrahedra $\bS_{cv}$ which form an interconnected two-dimensional network. This structure is topologically identical (albeit on a closed surface) to the networks of triangles described above in planar systems. In particular, the triangles on this surface also enclose polygonal facets. 

The surface of the cell can now be tiled in exactly the same way that we did planar structures. We define the centroids of the non-triangular polygonal facets on the cell surface as the mean position of their corners and name such a centroid $\zeta_p$.
We extend three lines from the centroid of every triangular facet, $\zeta_{cv}$, to the triangle corners (see figure 4). From $\zeta_{cv}$ to $\zeta_p$ we extend a vector $\bR_{cvp}$. This vector crosses a particular edge, from the ends of which we extend two lines to $\zeta_p$. This construction is shown in figure 4b. $\bR_{cvp}$ and the edge of the triangle that it crosses define a {\it non-planar (skew) quadrilateral}, an example of which is shown shaded in figure 4d. This and similar quadrilaterals tile the surface of the cell without overlap, leaving no gaps, exactly as in two-dimensional systems. The only difference here is that the two parts of the quadrilateral, the one within the triangular face and the one within the polygonal facet (shaded blue in figure 4), lie in planes that are tilted to one another due to the curvature of the surface. Such a quadrilateral can be uniquely indexed by the cell, $c$, on whose surface it resides, by the polygonal facet, $p$, on the surface and by the vertex, $v$, that gives rise to the original triangle $\bS_{cv}$. 

If we now extend straight lines from the four corners of such a quadrilateral to both the centroid of the cell, $\brho_c$ and to the centroid of vertex $\brho_v$, we get a non-convex octahedron (see figure 4f). The faces of this octahedron are shared with the faces of similar octahedra constructed on nearby quadrilaterals. The volume of the octahedron can be expressed in terms of the vectors $\bR_{cvp}$, $\bR_{cv}$, and the appropriate edge of the triangle  $\bS_{cv}$, which we term $\br_{cvp}$.
These octahedra {\it cover the entire three-dimensional space} without ovelaps, 

\begin{equation}
V_{sys} = \sum_{cvp} V_{cvp} \ .
\label{eq:Di}
\end{equation}

Thus, we have arrived at the first goal - the partition of the volume of a three-dimensional cellular structure into volumes of elementary building blocks; the octahedra. These are the quadrons of the three-dimensional structure for the purpose of the entropic analysis $V_{cvp}\to V_q$.

Let us count the total number of quarons. A tetrahedron surrounding a vertex has four triangular faces, each giving rise to three quadrilaterals. Thus, with each vertex we can associate a unique volume, composed of the volumes of the twelve three-dimensional octagonal quadrons, 

\begin{figure*}
\resizebox{1.0\textwidth}{!}
{%
  \includegraphics{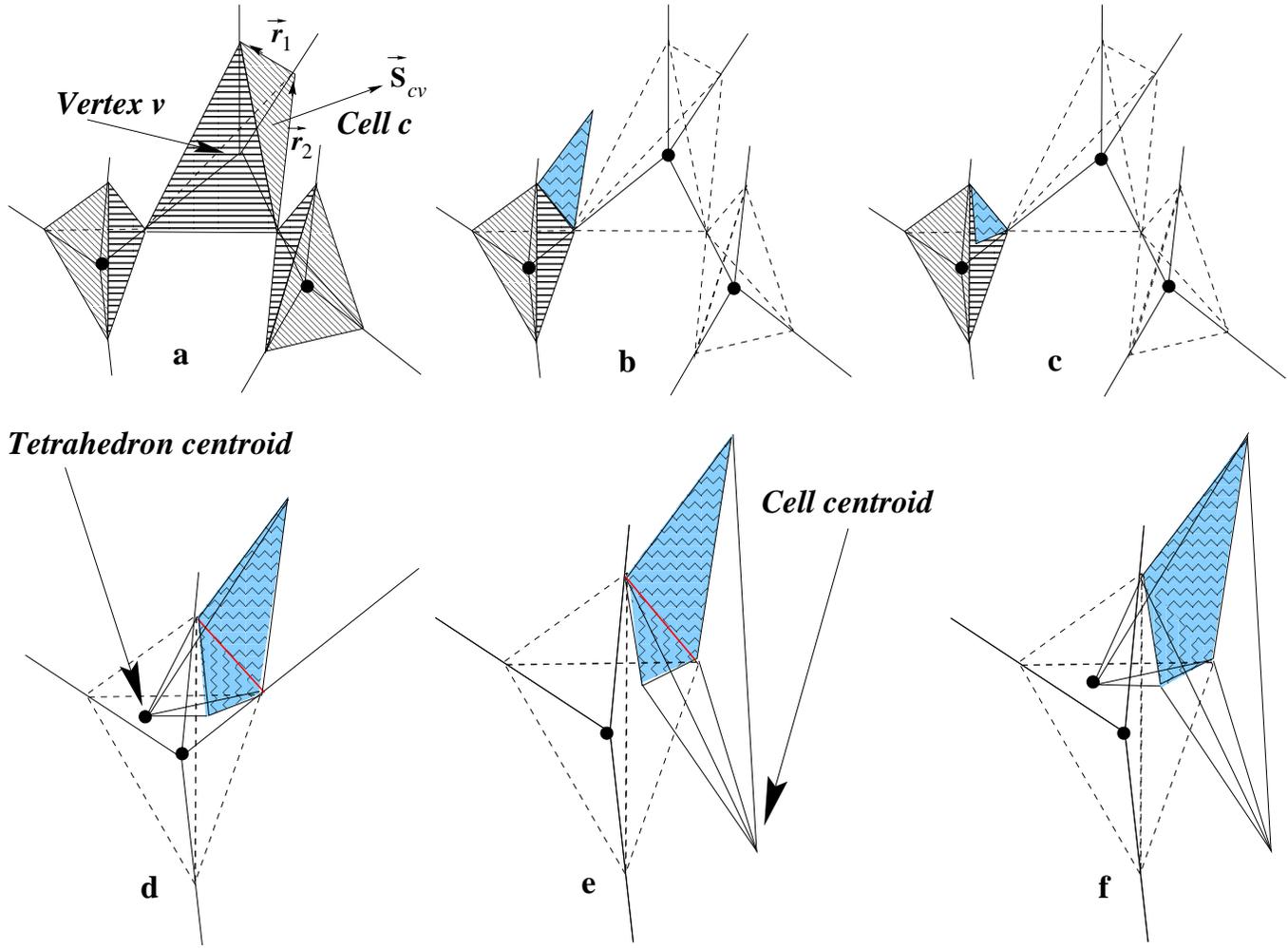}
}
\caption{The construction of a quadron: (a) Three tetrahedra on the boundary of one cell. (b,c) The two stages of the construction of a non-planar (skew) quadrilateral on the surface of the cell; First extend a triangle into the polygonal facet between triangles and then extend a triangle to the centre of the triangle $S_{cv}$. The diagonal of the skew quadrilateral (red line in 4d and 4e) is a section of the line where the planes of the polygon facet and $S_{cv}$ intersect. (d) Connecting the four corners of the quadrilateral to the centroid of tetrahedron $v$. (e) Connecting the four corners of the quadrilateral to the centroid of the cell. (f) The quadron is a non-convex octahedron extending between the centroids of the vertex and of the cell. There are twelve such octahedra around vertex $v$ whose joint volume makes a non-convex (stellated) icositetrahedron (i.e. of 24 faces). This is the volume associated with vertex $v$. These icositetrahedra perfectly fill the three-dimensional system and can also be used to partition its entire volume.}
\label{fig:4}       
\end{figure*}

\begin{equation}
V_v = \sum_{cp} V_{cvp} \ .
\label{eq:Dia}
\end{equation}
The twelve quadrons around every vertex make a star-like (stellated) icositetrahedron (i.e. a polyhedron of 24 faces) around it, with six faces extending into each of the neighbouring cells of the vertex.  We could then view the volume of the system as covered by the $N$ vertex icositetrahedra.

Next, let us determine the dimensionality of the configurational phase space. As in two-dimensions, correlations can only originate from vectors that close irreducible loops and our task is therefore to count these. The irreducible loops originate from all the polygonal facets on cell surfaces, including the triangles. All other loops can be expressed in terms of these. Consider then the surface of an arbitrary cell $c$, consisting of $n_v^c$ vertices and $n_f^c$ facets. Because on the cell surface the vertices are trivalent (the fourth strut is part only of neighbouring cells), the surface contains a total of $n_t^c = 3n_v^c/2$ struts. Each one of these struts can be described by a three-dimensional vector and so its specification requires three degrees of freedom. Every vertex is quadrivalent and every strut ends in two vertices, which gives that there are altogether $2N$ struts in the system. The $n_f^c$ irreducible loops on the cell surface give $n_f^c$ dependent strut vectors on this surface, which in turn gives that on this surface there are $3(3n_v^c/2-n_f^c)$ independent degrees of freedom. However, each facet is common to two cells and therefore, when summing this quantity over cells, every facet is exactly double-counted. This gives that the total number of dependent vectors is $\sum_c n_f^c /2$. Subtracting this number from the total number of struts, $2N$, we obtain that the total number of independent degrees of freedom is 

\begin{equation}
N_{dof} = 3(2N - \sum_c n_f^c /2) \ .
\label{eq:Dii}
\end{equation}  
This sum can be evaluated using Euler's relation for the topology of the surface of a cell, 

\begin{equation}
n_v^c - n_t^c + n_f^c = 2 \ ,
\label{eq:Diii}
\end{equation}
together with the observations that in quadrivalent open-cell structures \cite{BlFoamBook}

\begin{equation}
\sum_c n_v^c = 4N \ \ \ {\rm and}\ \ \ \sum_c n_t^c  = 6N \ .
\label{eq:Div}
\end{equation}
Combining these relations gives finally that the number of degrees of freedom is  

\begin{equation}
N_{dof} = 3\left[ 2N - \frac{1}{2}\sum_c ( 2 + n_t^c - n_v^c) \right] = 3(N-N_c) \ .
\label{eq:Dv}
\end{equation}
Unlike in two-dimensions, it is not possible to simplify this result further to obtain $N_{dof}$ in terms of $N$ alone; Euler's relation and the condition of quadrivalency are not sufficient for this. Nevertheless, relation (\ref{eq:Dv}) allows us to make two important observations. One is that $N_{dof} < 12N$ and therefore $N_{dof}$ is {\it smaller than the number of quadrons}. This means that, unlike in two dimensions, the quadrons are inter-correlated, which has significant implications for the construction of the partition function. 
The second observation is that $N_{dof} < 3N$, indicating that contrary to naive intuition, the dimensionality of the phase space of three-dimensional cellular systems is {\it even smaller} than that of two-dimensional systems. 

The partition function can now be written as 

\begin{equation}
Z = \int \cP(\{V_q\}) \prod_{q=1}^{N_{dof}} \left( e^{-\beta V_q } dV_q \right) \ ,
\label{eq:Dvi}
\end{equation}
where $\cP(\{V_q\})$ is the joint probability density of the volumes $V_q = V_{cvp}$ that are included explicitly in the integral. This probability density - the density of states - can be obtained from experimental measurements, from simulations, or from analytic calculations on case-study systems. Armed with the form of this function it is possible to compute the entropy, the mean porosity, porosity fluctuations and other related properties, in exactly the same way as was outlined for two-dimensional systems following eq. (\ref{eq:Ci}). 

Here too a first approximation would be to assume that the $N_{dof}$ quadrons are uncorrelated. This is the analogue of the ideal gas approximation and it simplifies the computation of  the partition function. To the best of our knowledge there exists currently no data in the literature on the form of the density of states $\cP(\{V_q\})$ in any system, either experimentally or analytically. Therefore, for illustration purposes, let us first assume that the quadron volumes have a Gaussian density of states,

\begin{equation}
\ccP(V_q) = C e^{-\frac{(V_q-V_0)^2}{2a^2}} \ .
\label{eq:Ei}
\end{equation}
In this expression $C$ is a normalisation constant and $V_q$ fluctuates around a mean value $V_0$ with a spread of $a$. If this is the steady-state distribution of dynamically evolving systems then $V_0$ and $a$ would scale with time. Substituting this form into the partition function we have

\begin{equation}
Z = \left[ \int_0^\infty Ce^{-\beta V_q - \frac{(V_q-V_0)^2}{2a^2}} dV_q \right]^{N_{dof}} = e^{N_{dof}(\frac{\beta^2 a^2}{2} - \beta V_0)} \ .
\label{eq:Eii}
\end{equation}
From this partition function we find that the entropy is 
\begin{equation}
S = ln Z - \beta\frac{\partial ln Z}{\partial \beta} = \beta^2 \frac{3(N - N_c) a^2}{2} \ ,
\label{eq:Eiii}
\end{equation}
the mean volume (or porosity) per quadron is 
$$\langle V \rangle = (V_0 - \beta a^2) \ ,$$
and the mean porosity fluctuations is
$$\langle \delta V^2 \rangle = [((V_0 - \beta a^2)^2 + a^2)] \ .$$

For our  second example we are inspired by data on polymeric open-cell foams. In these materials the cell size distribution is generically skewed  to the left \cite{Eletal02}. We are not aware of any fit to this distribution and therefore let us assume that it can be described by the form

\begin{equation}
\ccP(V_q) = C V_q^{x-1} e^{-V_q/V_0} \ \ \ ; \ \ \ 0 < V_q < V_{max} \ ,
\label{eq:Eiv}
\end{equation}
with $0<x<1$, $0 < V_{max}$, $V_0$ a typical size and $C=V_0^{-x}/\gamma(x,V_{max}/V_0)$ a normalisation constant. For $V_{max}\to\infty$ $\gamma(x,V_{max}/V_0) \to \Gamma(x)$. 
It is plausible that the quadron volume distribution follows the same distribution as the cell sizes and, using the density of states (\ref{eq:Eiv}), the partition function is

\begin{eqnarray}
Z & = & \left[ \int_0^{V_{max}} C V_q^{x-1} e^{-V_q/V_0} dV_q \right]^{N_{dof}} \nonumber \\
  & = & \left[ \frac{\left(\gamma(x,V_{max}(\beta + 1/V_0)\right)}{(V_0\beta + 1)^x \gamma\left( x,V_{max}/V_0 \right)} \right]^{N_{dof}} \ .
\label{eq:Ev}
\end{eqnarray}
Using now $z=Z^{1/N_{dof}}$ we can readily compute the entropy per quadron $s={\rm ln}z - \partial{\rm ln}z/\partial\beta$, the mean porosity per quadron $\langle V\rangle = - \partial{\rm ln}z/\partial\beta$, and the mean porosity fluctuations per quadron 
$\langle \delta V^2 \rangle = \partial^2{\rm ln}z/\partial\beta^2$. 

The form given in (\ref{eq:Dvi}) may not be the most convenient form to represent the partition function. In some situations it may be more convenient to use the more basic $3(N-N_c)$-dimensional phase space of the components of the independent vectors. For example, when the aim is to compute expectation values of structural properties that cannot be expressed directly in terms of quadron volumes. Such properties could be the mean area of polygonal facets on surface of cells, the mean moment of inertia of cells, or the anisotropy of the cellular structure. In this case the volumes of the quadrons (or the vertices, for that matter) could be first expressed in terms of the vectors that construct them, these expressions would be substituted into the volume function $\cW_3$, and finally a partition function of the form (\ref{eq:Aiv}) could be written down and computed.

\section{Discussion}
\label{sec:3}

To conclude, we have developed an exact characterisation of the configurational entropy of cellular structures in two and three dimensions. The two-dimensional formulation has been used to gain insight into the three-dimensional case. The analysis uses the concept of compactivity and we have constructed volume functions, which have several advantages: (i) They are exact in the sense that a linear sum over elementary entities of the structures gives the volume of the system. (ii) The analysis shows that the elementary entities are not vertices but rather quadrons, of which vertices are composed. In two dimensions and in three dimensions a vertex is composed of three and twelve quadrons, respectively. (iii) The volume functions make it possible to identify the basic variables on which the volumes depend. (iv) The functions make it possible to determine the correlations between variables and so identify the phase space of independent variables. We have found that in structure of $N$ vertices and $N_c$ cells the phase spaces in two and three dimensions are $3N$ and $3(N-N_c)$, respectively. (v) The volume function in two dimensions is derived directly from a fabric tensor which is also instrumental for stress analysis in such structures, had they been in mechanical equilibrium. This may provide a a bridge to a comprehensive theory of both stress and entropy of cellular solids and solid foams. 

It is interesting that in two dimensions the dimensionality of the phase space is the same as the number of quadrons, while in three dimensions it is much smaller. Moreover, we have found that the dimensionality of the phase space in three dimensions is smaller than that in two dimensions. The reason for this seemingly strange result is that, although the total number of variables in three-dimensional systems is much larger, so is the number of their inter-correlations. Therefore, the difference between these two quantities, which gives the dimensionality of the phase space, does not necessarily increase with the dimension in which the physical system is embedded. 

Several case studies of densities of states, i.e.  distributions of quadron volumes, have been analysed. For these cases we have calculated explicitly the mean porosity and the mean porosity fluctuations. The entropy has been calculated explicitly for one case in two dimensions and one case in three and its calculation has been outlined for the other cases.
Since we have only aimed to illustrate the method, all the examples that we have studied contain no correlations between quadrons. It is not difficult to study such correlations for model systems - an example of such a calculation in two dimensions has been carried out in reference \cite{BlEd03}, albeit for a granular system (but it applies equally well to cellular structures).
Writing the partition function in terms of the density of states $\cP$ eliminates any explicit dependence on dimensionality, making the computation of the entropy in two- and three dimensions formally identical. Dimensional information is then contained implicitly both in the form of $\cP$ and in the number of quadrons. 
For ideal binary mixtures of quadrons we have derived an analogue of the Bragg-Williams description for alloys. 
The three-dimensional systems that we have presented are idealised in that we have assumed no correlations between quadrons and therefore may be regarded as an `ideal quadron gas approximation'.
Note that the number of independent quadrons is $3(N-N_c)$ - a small fraction ($<1/4$) of the $12N$ quadrons in the system. Choosing these to be uniformaly distributed throughout the system means that their spatial density is quite low. The rest of the quadrons may be regarded as {\it frozen degrees of freedom}. 
It is evident that the entropy depends sensitively on the density of states $\cP$ but, to our knowledge, there is no derivation in the literature of its form in any system, nor is there experimental or numerical data on it. However, several works on cellular systems have analysed cell size distributions and by adopting these to quadron size distributions it may be possible to make progress.  This is the route we have taken in our second example in three-dimensions.


\begin{thebibliography}{99}
\bibitem{irreducible} These are the irreducible loops, i.e., the smallest set of loops in terms of which all other loops can be expressed
\end{thebibliography}
\end{document}